\documentclass[preprint,showkeys]{revtex4}

\usepackage{graphics}
\usepackage{graphicx}
\usepackage{amssymb}
\usepackage{amsmath}
\usepackage{amsfonts}
\usepackage{setspace}

\newif\ifFIG
\FIGtrue

\begin{document}

%\begin{titlepage}
%\begin{center}

%{\Large
% \bf{Theoretical study of lithium clusters by electronic stress tensor}
% }

%\vskip .45in

%{\large
%Kazuhide Ichikawa, Hiroo Nozaki, Naoya Komazawa \\ and Akitomo Tachibana$^{*}$}

%\vskip .45in

%{\em
%Department of Micro Engineering, Kyoto University, 
%Kyoto, 606-8501, Japan
%}

%\vskip .45in
%{\tt E-mail: akitomo@scl.kyoto-u.ac.jp}

%\end{center}

%\vskip .4in

%\date{\today}

%{Wave function analysis; Theory of chemical bond; Stress tensor; Lithium clusters; Metallic bond}

\title{Theoretical study of lithium clusters by electronic stress tensor}
\author{Kazuhide Ichikawa}
\author{Hiroo Nozaki}
\author{Naoya Komazawa}
\author{Akitomo Tachibana}
\email{akitomo@scl.kyoto-u.ac.jp}
%\address{Department of Micro Engineering, Kyoto University, Kyoto, 606-8501, Japan}
\affiliation{Department of Micro Engineering, Kyoto University, Kyoto, 606-8501, Japan}

\begin{abstract}
We study the electronic structure of small lithium clusters Li$_n$ ($n=2\sim 8$) using the electronic stress tensor. 
We find that the three eigenvalues of the electronic stress tensor of the Li clusters are negative and degenerate,
just like the stress tensor of liquid. 
This leads us to propose that we may characterize a metallic bond in terms of the electronic stress tensor. 
Our proposal is that in addition to the negativity of the three eigenvalues of the electronic stress tensor, 
their degeneracy characterizes some aspects of the metallic nature of chemical bonding.
To quantify the degree of degeneracy, we use the differential eigenvalues of the electronic stress tensor.
By comparing the Li clusters and hydrocarbon molecules, we show that the sign of the largest eigenvalue
and the differential eigenvalues could be useful indices to evaluate the metallicity or covalency of a chemical bond. 
\end{abstract}

\keywords{Wave function analysis; Theory of chemical bond; Stress tensor; Lithium clusters; Metallic bond}

\maketitle

%\end{titlepage}

%\setcounter{page}{1}

%\tableofcontents

%%%%%%%%%%%%%%%%%%%%%%%%%%%
\section{Introduction} \label{sec:intro}
%%%%%%%%%%%%%%%%%%%%%%%%%%%

The stress tensors in general are widely used for description of internal forces of matter in various fields of science such as mechanical engineering and material science. 
In quantum systems as well, the stress tensors have been investigated for many years, including one of the earliest quantum mechanics papers 
\cite{Schrodinger1927, Pauli, Epstein1975, Bader1980, Bamzai1981a, Nielsen1983, Nielsen1985, Folland1986a, Folland1986b, Godfrey1988, Filippetti2000, Tachibana2001, Pendas2002, Rogers2002, Tachibana2004, Tachibana2005, Morante2006, Tao2008, Ayers2009, Tachibana2010,Jenkins2011,GuevaraGarcia2011,Tachibana2012}. 
As for the stress tensors in quantum mechanics context, we can find several different definitions and applications in the literature.
For example, Ref.~\cite{Nielsen1983} and followers focus on the stress tensor which is associated with forces on nuclei. 
In contrast, the one we consider in this paper is the electronic stress tensor, which is associated with effects caused by internal forces acting on electrons in molecules, following Ref.~\cite{Tachibana2001}. 
This electronic stress tensor has been used to investigate chemical bonds and reactions and many interesting properties have been discovered 
\cite{Tachibana2001,Tachibana2004,Tachibana2005, Szarek2007, Szarek2008, Szarek2009, Ichikawa2009a, Ichikawa2009b,Tachibana2010, Ichikawa2010, Ichikawa2011a, Ichikawa2011b, Ichikawa2011c}. 

We briefly review our past works on the electronic stress tensor which are closely related to this paper. 
In Ref.~\cite{Tachibana2004}, it has been proposed that a covalent bond can be described by the eigenvalues and
 eigenvectors of the electronic stress tensor. 
In detail, the bonding region with covalency can be characterized and visualized by the ``spindle structure",
where the largest eigenvalue of the electronic stress tensor is positive and
 the corresponding eigenvectors form a bundle of flow lines that connects nuclei. 
In Ref.~\cite{Szarek2007}, the eigenvalues of the electronic stress tensor of various homonuclear diatomic molecules 
have been investigated and all metals and metalloids molecules studied there exhibit
the negative largest eigenvalue at the midpoint of the nuclei. 
Ref.~\cite{Szarek2007} has mentioned that this negativity of the largest eigenvalue may be connected to the metallic nature
of bonding but, at the same time, it has been known that the negative largest eigenvalue appears between
a pair of atomic nuclei with very short distance such as the C-C bond in C$_2$H$_2$ \cite{Tachibana2005,Szarek2007}.
In Refs.~\cite{Ichikawa2011a} and \cite{Ichikawa2011c}, the Al$_4$ and Pd clusters have been investigated 
and we have shown ``pseudo-spindle structure" \cite{Ichikawa2011c} between metal nuclei.
The pseudo-spindle structure is a negative eigenvalue version of the spindle structure and, more precisely, 
it is the region between two atoms where the largest eigenvalue of the
electronic stress tensor is negative and corresponding eigenvectors forming a pattern
which connects them.
However, again, the pseudo-spindle structure is also seen in C$_2$H$_2$ \cite{Tachibana2005,Ichikawa2011b}.
Since the pseudo-spindle structure is seen in many bonds between metal nuclei, it is tempting to associate it 
with a metallic bond, but it seems that the pseudo-spindle structure is not sufficient to
capture the entire feature of a metallic bond.

In this paper, we would like to argue how we may describe a metallic bond using the electronic stress tensor. 
The conventional wisdom would say that it is not sensible to define ``a metallic bond" because metal is something 
defined only for bulk with many electrons and ions. 
Our challenge is capturing some aspects of a metallic bond or metallicity of a chemical bond using the electronic stress tensor.
As a starting point, we tackle this problem using small clusters of lithium, the simplest metal.

There are many studies on the nature of bonding in the Li clusters in the literature using such methods as
by generalized valence-bond \cite{McAdon1985}, 
electron localization function (ELF) \cite{Rousseau2000,Alikhani2006} and 
quantum theory of atoms in molecules (QTAIM) \cite{Gatti1987,Bersuker1993,Yepes2012}.
We would also like to see whether our stress tensor approach gives complementary views to these methods.

The structure of this paper is as follows. 
Sec.~\ref{sec:calc} briefly summarizes our analysis method of electronic structures using the electronic stress tensor.
Sec.~\ref{sec:results} shows our results of stress tensor analysis on the Li clusters. 
In Sec.~\ref{sec:dinuclear}, we discuss Li$_2$. We also discuss H$_2$ and LiH for comparison.
In Sec.~\ref{sec:clusters}, the Li clusters are analyzed and a method to use differential eigenvalues of the stress tensor is proposed.
In Sec.~\ref{sec:comparison}, that method is  applied to hydrocarbon molecules and its effectiveness is tested.
Finally, Sec.~\ref{sec:conclusion} is devoted to our conclusion. 

%%%%%%%%%%%%%%%%%%%%%%%%%%%%%%%%%%%
\section{Theory and calculation methods} \label{sec:calc}
%%%%%%%%%%%%%%%%%%%%%%%%%%%%%%%%%%%

In this section, we summarize quantities and method we use for our electronic stress tensor analysis.
Our stress tensor is based on the rigged quantum electrodynamics (QED) theory \cite{Tachibana2001}.
The rigged QED is the ordinary QED (the theory of photons and electrons) equipped with degrees of freedom of atomic nuclei
 in order to be used for atomic or molecular systems.
Since this is a theory of quantum fields, there exist quantum operators defined at each point in spacetime. 
By using the equation of motion for operators of various physical quantities, many interesting operator relations have been found 
\cite{Tachibana2001,Tachibana2004,Tachibana2005,Tachibana2010}.

The one related to the electronic stress tensor is the equation of motion of the kinetic momentum density operator.
It turns out that the time derivative of the kinetic momentum density operator can be expressed by the sum of the two force operators:
the tension density operator $\Hat{\vec{\tau}}^S(\vec{r})$ and the Lorentz force density operator $\Hat{\vec{L}}(\vec{r})$. 
$\Hat{\vec{\tau}}^S(\vec{r})$ can be expressed by the divergence of some 3 by 3 tensor operator, which we call
 the stress tensor density operator $\hat{\overleftrightarrow{\tau}}^{S}(\vec{r})$. 
We obtain the quantum field theory version of the equilibrium equation of stress tensor 
by taking the expectation value of this equation of motion with respect to stationary electronic state of an atom or molecule
(note that the expectation value of the time derivative of any operator with respect to stationary state is zero) as
$0 = \vec{\tau}^S(\vec{r}) + \langle \Hat{\vec{L}}(\vec{r}) \rangle$ or 
$0 = \sum_l \partial_l  \tau^{Skl}(\vec{r}) + \langle \Hat{L}^k(\vec{r})\rangle$
where $\{k, l\} = \{1, 2, 3\}$, 
$\langle ... \rangle$ denotes the expectation value,
$\vec{\tau}^S(\vec{r}) \equiv \langle \hat{\vec{\tau}}^S(\vec{r}) \rangle$ and
$\tau^{Skl}(\vec{r}) \equiv \langle \hat{\tau}^{Skl}(\vec{r})  \rangle$

The expectation values for the stress tensor density and tension density can be expressed as
\begin{eqnarray} 
\tau^{Skl}(\vec{r}) &=& \frac{\hbar^2}{4m}\sum_i \nu_i
\Bigg[\psi^*_i(\vec{r})\frac{\partial^2\psi_i(\vec{r})}{\partial x^k \partial x^l}-\frac{\partial\psi^*_i(\vec{r})}{\partial x^k} \frac{\partial\psi_i(\vec{r})}{\partial x^l} \nonumber\\
& & \hspace{4cm} +\frac{\partial^2 \psi^*_i(\vec{r})}{\partial x^k \partial x^l}\psi_i(\vec{r}) -\frac{\partial \psi^*_i(\vec{r})}{\partial x^l}\frac{\partial \psi_i(\vec{r})}{\partial x^k}\Bigg], \label{eq:stress} \\
\tau^{S k}(\vec{r})&=&  \sum_l \partial_l  \tau^{Skl}(\vec{r}) \nonumber \\
&=&\frac{\hbar^2}{4m}\sum_i \nu_i
\Bigg[\psi^*_i(\vec{r})\frac{\partial \Delta\psi_i(\vec{r})}{\partial x^k}-\frac{\partial\psi^*_i(\vec{r})}{\partial x^k} \Delta\psi_i(\vec{r}) \nonumber\\
& & \hspace{4cm} +\frac{\partial \Delta\psi^*_i(\vec{r})}{\partial x^k}\psi_i(\vec{r}) -\Delta \psi^*_i(\vec{r}) \frac{\partial \psi_i(\vec{r})}{\partial x^k}\Bigg],
\label{eq:tension}
\end{eqnarray}
where $\{k, l\} = \{1, 2, 3\}$, $m$ is the electron mass, $\psi_i(\vec{r})$ is the $i$th natural orbital and $\nu_i$ is its occupation number.
Since the stress tensor is an Hermitian matrix, its eigenvalues are real and eigenvectors are orthogonal to each other. 
We denote the eigenvalues by $\lambda_i$ $(i=1,2,3)$ with the ordering of $\lambda_3 \ge \lambda_2 \ge \lambda_1$.

The concept of  ``Lagrange point"  is proposed in Ref.~\cite{Szarek2007} and we use it as a point which can characterize 
a bond between two atoms.  
 The Lagrange point $\vec{r}_L$ is the point where the tension density $\vec{\tau}^S(\vec{r})$
vanishes: $\tau^{S k}(\vec{r}_L)=0$. 
We report the eigenvalues of electronic stress tensor at this point in the following sections.

We also analyze electronic structure of molecules using kinetic energy density $n_T(\vec{r})$, 
\begin{eqnarray}
n_{T}(\vec{r}) = - \frac{\hbar^2}{4m}
\sum_{i} \nu_i \left[ \psi_{i}^{*}(\vec{r}) \Delta \psi_{i}(\vec{r}) + 
\Delta \psi_{i}^{*}(\vec{r}) \cdot \psi_{i}(\vec{r}) \right],   \label{eq:KED} 
\end{eqnarray}
which is defined in Ref.~\cite{Tachibana2001}.
Note that our definition of the kinetic energy density is not positive-definite. 
Using this kinetic energy density, we can divide the whole space into three types of region:
 the electronic drop region $R_D$ with $n_T(\vec{r}) > 0$, 
where classically allowed motion of electron is guaranteed and the electron density is amply accumulated;
the electronic atmosphere region $R_A$ with $n_T(\vec{r}) < 0$, where the motion of electron is classically forbidden
and the electron density is dried up;
and the electronic interface $S$ with $n_T(\vec{r}) = 0$, the boundary between $R_D$ and $R_A$, which corresponds to a turning point. 
 The $S$ can give a clear image of the intrinsic shape of atoms and molecules and is therefore an important region in particular. 

The electronic structures used in this paper are obtained by the Gaussian 09 \cite{Gaussian09} for cluster models 
and ABINIT \cite{ABINIT1,ABINIT2} for periodic models.
We use Molecular Regional Density Functional Theory (MRDFT) package \cite{MRDFTv3} to compute the quantities, Eqs.~\eqref{eq:stress}-\eqref{eq:KED},
 introduced in this section. Some part of the visualization is done using PyMOL Molecular Viewer program \cite{PyMOL}.

%%%%%%%%%%%%%%%%%%%%%%%%%%%%%%%
\section{Results and discussion} \label{sec:results}
%%%%%%%%%%%%%%%%%%%%%%%%%%%%%%%

%%%%%%%%%%%%%%%%%%%%%%%%%%%%%%%
\subsection{Li$_2$} \label{sec:dinuclear}
%%%%%%%%%%%%%%%%%%%%%%%%%%%%%%%

In this section, we analyze the smallest Li cluster, Li$_2$, by the electronic stress tensor.
For comparison, we also show the results of stress tensor analysis of H$_2$ and LiH.
The computation is performed by the coupled-cluster single-double (CCSD) method using the 6-311++G** basis set \cite{Raghavachari80b,Frisch84}.

Before we discuss Li$_2$, we review how the chemical bond of the hydrogen molecule is expressed by the electronic stress tensor
 (Eq.~\eqref{eq:stress}). 
 In Fig.~\ref{fig:dist_H2}~(b), we plot the largest eigenvalue of the stress tensor and corresponding eigenvector on the plane including 
 the internuclear axis. 
If the largest eigenvalue of the stress tensor is positive, it is called the tensile stress,
 and if negative, it is called the compressive stress.
Namely, the sign of the largest eigenvalue tells whether electrons at a certain point in space 
exhibit tensile stress or compressive stress.
The difference between tensile stress and compressive stress are described as follows.
If we consider a fictitious plane at some point in atomic or molecular system, there are two ways that 
a part on one side of the plane acts on the other part of the plane.
One is when electrons on one side of the plane is ``pulled up" by the electron field on the other side.
The other is when electrons on one side is ``pushed back" by the electron field on the other side.
The former corresponds to the tensile stress and the latter to the compressive stress.
The direction of the eigenvector corresponds to the direction normal to the plane.
 In Fig.~\ref{fig:dist_H2}~(b), we can see that the region with positive eigenvalue spreads between the H atoms and,
in that region, the eigenvectors form a bundle of flow lines that connects the H nuclei. 
Such a region is called ``spindle structure" \cite{Tachibana2004} and it is clearly seen in the figure.
This means that electrons close to one H nucleus are pulled up by the electron field close to another H nucleus.
Such a structure can be associated with the formation of the tight bonding and has a suitable directionality
for a bond between two H nuclei.
Thus, it is proposed that such positive eigenvalue region is the manifestation of 
the strong covalent bond \cite{Tachibana2004}.

We now turn to Li$_2$, whose largest eigenvalue of the stress tensor and corresponding eigenvector
is shown in Fig.~\ref{fig:dist_Li2}~(b) in the same manner as H$_2$.
The striking difference is that there is a region with the negative largest eigenvalue between Li nuclei and there is no spindle structure. 
We do not even see a ``pseudo-spindle structure" \cite{Ichikawa2011c}, the negative eigenvalue version of the spindle structure
({\it i.e.} the region with the eigenvectors forming a bundle of flow lines that connects two nuclei but with negative eigenvalues),
which is frequently seen between metallic atoms \cite{Ichikawa2011a, Ichikawa2011b, Ichikawa2011c}.
This means that the electron field pushes back electrons in the neighboring regions with no solid directionality which can be
considered as a bond axis.
This is similar to the stress tensor of liquid.
 A liquid has the ability to flow. 
 Liquid particles are loosely bound and they can move around one another freely.
Such properties can be well associated with electrons in metal.
In fact, coincidentally, metallic bonding is often described as  
an ``electron sea"  and the positively charged metal ions.
It may be unusual to say Li$_2$ has purely a metallic bond with no trace of covalency
but if we look at the electronic stress tensor, it is totally different from a covalent bond 
and certainly has a feature which can be called metallic.
We will return to this point when we discuss larger Li clusters. 

After seeing the difference between H$_2$ and Li$_2$ in terms of the electronic stress tensor, 
it may be interesting to see what the stress tensor of LiH look like.
This is shown in Fig.~\ref{fig:dist_LiH}~(b).
It has both positive and negative regions between H and Li nuclei. 
The region close to H/Li has positive/negative eigenvalue.
In the region very close to the internuclear axis, there is some flow of eigenvectors 
in the direction of connecting the nuclei, but it is somewhat different from spindle or pseudo-spindle structures.
Therefore, we can say that LiH has a bond with solid directionality like covalent bond of H$_2$ 
but there is a strong part nearby H and a weaker part nearby Li. 
It is tempting to associate this with an ionic bond, but we reserve conclusion until 
more examples are examined. 

Since we have discussed the analogy between the stress tensors of Li$_2$ and liquid, it is important to investigate 
the other two eigenvalues.
We plot all the three eigenvalues of the stress tensor for H$_2$, Li$_2$ and LiH along the internuclear axes in Fig.~\ref{fig:eig_Li2H2LiH}.
The bond length and the eigenvalues at the Lagrange point are summarized in Table \ref{tab:Li2H2LiH}.
Let us first look at H$_2$. 
As shown in Fig.~\ref{fig:eig_Li2H2LiH} (a), around the midpoint of the two H nuclei, 
the largest eigenvalue ($\lambda_3$) is positive and the smaller eigenvalues ($\lambda_1$ and $\lambda_2$) 
are negative and degenerate.
The absolute values of the three eigenvalues are ${\cal O}(0.1)$ and so are the differences between $\lambda_3$ and $\lambda_1$ or $\lambda_2$.
Fig.~\ref{fig:eig_Li2H2LiH} (b) shows the case of Li$_2$ and we see that the three eigenvalues are all negative
around the midpoint of the two Li nuclei.
Two larger eigenvalues ($\lambda_3$ and $\lambda_2$) are degenerate and the smallest one ($\lambda_1$)
is smaller than them by only ${\cal O}(10^{-4})$.
Therefore, compared to the case of H$_2$, the three eigenvalues of Li$_2$ can be regarded as degenerate. 
Such degeneracy in the eigenvalues of stress tensor, suggesting the lack of directionality, solidifies the analogy with liquid. 
Note that, in terms of the stress tensor, liquid is defined by the three eigenvalues which are negative and degenerate.
This is in stark contrast with the case of H$_2$, whose eigenvalue pattern indicates the strong directionality of the bond. 
The case of LiH is shown in Fig.~\ref{fig:eig_Li2H2LiH} (c). 
The three eigenvalues are almost degenerate in the region close to Li and the region close to H has positive largest eigenvalue
and two smaller eigenvalues are negative and degenerate. 
It is different from both Li$_2$ and H$_2$ and it looks like a mixture of Li$_2$ and H$_2$.
Such a feature could be a hint to consider how an ionic bond may be expressed by the electronic stress tensor,
but, again, we need more examples.

So far, we have analyzed the molecules at the equilibrium distances and discussed their nature of bonding.
We now try to extract more information by changing the internuclear distances.
In particular, if we compute the stress tensors at very long distances, we can study the nature of bonding
from the viewpoint of the bond formation.

The eigenvalues and eigenvectors of the stress tensor for H$_2$, Li$_2$ and LiH at various internuclear distances 
are plotted in Figs.~\ref{fig:dist_H2}, \ref{fig:dist_Li2} and \ref{fig:dist_LiH}.
In the following discussion, the zero surface of the kinetic energy density (Sec.~\ref{sec:calc}, Eq.~\eqref{eq:KED}),
the electronic interface $S$, plays an important role.
As is explained in Sec.~\ref{sec:calc}, $S$ is the boundary between 
 the electronic drop region $R_D$ (where the electron density is accumulated) 
 and the electronic atmosphere region $R_A$ (where the electron density is dried up).
In the figures, $S$ is drawn by green dashed lines. 
To identify which side of $S$ has positive or negative kinetic energy density ($R_D$ or $R_A$),
 it is useful to remember $R_D$'s exist, naturally, around nuclei. 
When we look at the figures with very long internuclear distances, we see $S$ exists around each nucleus,
indicating that they do not form molecules and suggesting the existence of independent atoms.
Moving to the figures with shorter internuclear distances, we see that $S$'s around nucleus get closer, merge
and eventually form $S$ for the molecules at the equilibrium distances. 
Incidentally, the state when $S$'s just begin to merge, namely touch each other, is called 
the ``intrinsic electronic transition state" \cite{Tachibana2001}. 
We denote the internuclear distance at this state by $r^{\rm TS}$.

The notable fact is that, when the distance is so long that $S$'s are well separated
(when practically considered as an infinite distance), we see the spindle structure at $R_A$ 
between the two nuclei for all of H$_2$, Li$_2$ and LiH
(see Figs.~\ref{fig:dist_H2} (d), \ref{fig:dist_Li2} (h) and \ref{fig:dist_LiH} (d) respectively).
This means that electrons in $R_D$ nearby nuclei are pulled up toward $R_A$ through $S$, resulting in the formation 
of a Lewis electron pair. 
In other words, all the molecules are covalently bonded at the very long distances. 
In the case of H$_2$, as shown in Fig.~\ref{fig:dist_H2}, the spindle structure continues to exist at shorter distances including $r^{\rm TS}_{\rm HH}=2.20\,{\rm \AA}$ and the equilibrium distance. 

In the case of Li$_2$, more interesting things happen. 
First, we observe that the spindle structure exist at the intrinsic electronic transition state 
($r^{\rm TS}_{\rm LiLi} = 5.43\,{\rm \AA}$, Fig.~\ref{fig:dist_Li2} (g)). 
Then, at 3.31\,${\rm \AA}$, the positive eigenvalue region vanishes and the spindle structure turns into 
the pseudo-spindle structure. 
(Note that, at slightly longer distance of 3.36\,${\rm \AA}$, the positive eigenvalue region vanishes on the 
internuclear axis while positive areas remain away from the axis. 
Namely, between 3.36\,${\rm \AA}$ and 3.31\,${\rm \AA}$, the positive region is not simply connected.)
That pseudo-spindle structure is broken at 2.78\,${\rm \AA}$, at which three eigenvalues are degenerate (Fig.~\ref{fig:dist_Li2eig} (b)).
At shorter distances, we see a negative region with no particular directionality such as to connect two nuclei, especially at the equilibrium distance.
This is an example of what has been argued in Ref.~\cite{Tachibana2012}.
When there are two distant Li atoms, the tensile stress (as is indicated by the spindle structure) pulls up electron 
in $R_D$ to the adjacent $R_A$ through $S$ which separates them. 
The consequence is the formation of the long-range Lewis pair of electron.
As the distant pair of Li atoms comes closer, the spindle structure disappears, indicating that the Lewis pair is unbound.
This is considered to be a manifestation of metallicity. 

In the case of LiH, at long distances, $R_A$ between Li and H is characterized by the spindle structure as mentioned above
and, in particular at $r^{\rm TS}_{\rm LiH} = 3.38\,{\rm \AA}$. 
In Fig.\ref{fig:dist_LiH} (c), we see the contact point of the two interfaces is covered by the spindle structure. 
When the distance gets shorter, although the spindle structure is partly lost, the positive eigenvalue region remains around H.
The negative eigenvalue region develops in the region close to Li but do not cover the whole space unlike the case of Li$_2$.
As for LiH, from the viewpoint of the electronic stress tensor, we may conclude that it has a bond which is neither covalent nor metallic. 
Whether we can categorize it as ionic or other bonding is an interesting question to ask and will be investigated in future. 

%%%%%%%%%%%%%%%%%%%%%%%%%%%%%%%
\subsection{Lithium clusters} \label{sec:clusters}
%%%%%%%%%%%%%%%%%%%%%%%%%%%%%%%
The optimized structures for Li clusters, Li$_n$ $(n=3\sim 8)$ are shown in Fig.~\ref{fig:structure_Licluster}. 
In the figure, atoms are connected when the Lagrange point (Sec.~\ref{sec:calc}) is found between them.
These structures are obtained by re-optimizing the structures reported in Ref.~\cite{Gardet1996}. 
We performed optimization with the three lowest multiplicities for each cluster and adopted the one with the lowest energy, 
which turned out to be singlet for clusters with even numbers of atoms and doublets for those with odd numbers. 
The computation is performed by the CCSD method using the 6-311++G** basis set \cite{Raghavachari80b,Frisch84}.
Although computational setups are slightly different, the obtained structures are basically consistent with Ref.~\cite{Gardet1996}. 
As for Li$_6$, Ref.~\cite{Gardet1996} has reported the structure with $D_{2h}$ symmetry while
we have obtained the one with higher symmetry of $D_{4h}$, which is consistent with Ref.~\cite{Florez2008}.

We first examine the largest eigenvalue of the stress tensor and corresponding eigenvector on the plane including 
three atoms in the Li clusters. 
They are plotted in Figs.~\ref{fig:eigvec_Licluster1} and \ref{fig:eigvec_Licluster2}.
We omit equivalent bonds due to the structural symmetry. 
In these figures, every pair of atoms has a Lagrange point in between. 
We focus on the electronic drop regions ({\it i.e.}~regions with positive kinetic energy density) between two atoms in each panel. 
The common feature is that they all have the largest eigenvalue which is negative.
As for the eigenvectors, many regions do not show clear pattern such as to connect two atoms, just as the case of Li$_2$. 
The regions with rather clear pseudo-spindle structures are found but not so common
(the 1-2 bond in Li$_6$ (Fig.~\ref{fig:eigvec_Licluster1} (f)),
the 1-2 bond in Li$_7$ (Fig.~\ref{fig:eigvec_Licluster2} (a)) and  
the 1-2 bond in Li$_8$ (Fig.~\ref{fig:eigvec_Licluster2} (c) or (d)).

From the discussion in Sec.~\ref{sec:dinuclear}, 
in order to characterize a chemical bond by the electronic stress tensor, 
it is important to know the pattern and degree of degeneracy of the three eigenvalues in addition to 
the sign of the largest eigenvalue.
The three eigenvalues of stress tensor at the Lagrange points are summarized in Table \ref{tab:Licluster}.
From the table, we can reconfirm that the eigenvalues are all negative.
In addition, we see that the three eigenvalues are close to each other as is the case of Li$_2$ at least at the Lagrange points.

In order to quantify the degeneracy pattern more clearly, we propose to use following differential eigenvalues.
\begin{eqnarray}
\lambda_{D32} &\equiv& \lambda_3-\lambda_2,    \\
\lambda_{D21} &\equiv& \lambda_2-\lambda_1.
\end{eqnarray}
Since our convention is $\lambda_3 \ge \lambda_2 \ge \lambda_1$, 
$\lambda_{D32}$ and $\lambda_{D21}$ are positive.
For example, at the Lagrange point, 
$(\lambda_{D32}, \lambda_{D21}) =$ $(0.394, 0)$ for H$_2$,
$(0, 2.37 \times 10^{-4})$ for Li$_2$
and $(8.52 \times 10^{-3}, 0)$ for LiH. 
In Fig.~\ref{fig:diff_eig} (a), we plot $(\lambda_{D32}, \lambda_{D21})$ at the Lagrange points for the Li clusters. 
It is shown that the Li clusters have very small values of $\lambda_{D32}$ and $\lambda_{D21}$, 
the same order as that of $\lambda_{D21}$ of Li$_2$ and much smaller than $\lambda_{D32}$ of H$_2$.
Thus, we can consider three eigenvalues of the Li clusters are almost degenerate,
meaning the high degree of isotropy (low degree of directionality) of the electronic stress tensor. 
Together with their negativity, the stress tensor of the Li clusters are just like that of liquid.
As we have mentioned in Sec.~\ref{sec:dinuclear}, electrons in metal are readily associated with particles in liquid.
By turning this argument around, we may try to consider that negativity and degeneracy of the eigenvalues of 
the electronic stress tensor characterize and quantify some aspects of the metallic nature of chemical bonding. 

To support the validity of this idea, we here show some data of very small Na clusters, Na$_n$ $(n=2\sim 4)$.
The electronic structures are obtained by the same computational setups as the Li clusters. 
We performed geometrical optimization starting from the structures reported in Refs.~\cite{Solovyov2002,Florez2008}.
The most stable structures turned out to be singlet for clusters with even numbers of atoms and doublets for those with odd numbers. 
As for the symmetry of those structures, Na$_3$ and Na$_4$ are found to have $C_{2v}$ and $D_{2h}$ symmetry.
The results of the electronic stress tensor analysis are summarized in Table \ref{tab:Nacluster}.
From the data in the table, we readily see that the eigenvalues are all negative and almost degenerate just as the case of Li clusters. 

As another way to test this idea, we analyze the electronic stress tensor of periodic models for bulk Li and Na.
The body-centered cubic structure is adopted for both Li and Na with the lattice constants taken to be
3.491\,{\AA} and 4.225\,{\AA}  respectively.
We use the norm-conserving pseudopotentials of Troullier-Martins type \cite{Troullier1991} and 
the generalized-gradient approximation method by Perdew-Burke-Ernzerhof \cite{Perdew1996} for density
functional exchange-correlation interactions.
Kinetic energy cutoff of plane-wave expansion (k-point) is taken as 40.0 hartree ($2\times 2\times 2$ k-point set).
We compute the electronic stress tensor at the midpoint of two nearest neighborhood atoms.
% bond distance 3.023A for Li and 3.659 for Na
We obtain $\lambda_3 = -0.468 \times 10^{-3}$ and $\lambda_2 = \lambda_1 = -0.716 \times 10^{-3}$ for Li,
and $\lambda_3 = \lambda_2 = -0.253 \times 10^{-3}$ and $\lambda_1 = -0.293 \times 10^{-3}$ for Na.
As for the differential eigenvalues, 
$(\lambda_{D32}, \lambda_{D21}) =$ $(2.48 \times 10^{-4}, 0)$ for Li and $(0, 3.95 \times 10^{-5})$ for Na.
These results show that the negativity and degeneracy of the eigenvalues of the electronic stress tensor are found 
in the bulk systems too, and the degrees of negativity and degeneracy are not different from the clusters. 

Here, some comments are in order as for comparison with the studies in the literature. 
As shown just above, from the viewpoint of the electronic stress tensor, the chemical bonds in the small Li clusters 
look similar to the bulk Li. They posses compressive stress (negative eigenvalues) with no particular directionality 
(almost degenerate eigenvalues) like liquid, which can be considered as one of the important features of metal.
Interesting thing is that this is even so with Li$_2$.
Past studies like Refs.~\cite{McAdon1985,Rousseau2000} show multi-center bonds are ubiquitous
features of the Li clusters. In particular, Ref.~\cite{Rousseau2000} argues that 
the metallic phase of Li is characterized by multi-center bonds and very gentle variation of ELF across space.
From this viewpoint, since Li$_2$ is not multi-centered by definition, Li$_2$ does not have a common feature
with bulk Li or the other Li clusters. 
This point is in stark contrast with the stress tensor viewpoint. 
We note that, in Ref.~\cite{Gatti1987,Bersuker1993}, where Li$_2$ is studied in the framework of QTAIM,
Li$_2$ is viewed as a system that two Li atoms are bonded to a pseudo atom (the central nonnuclear attractor)
and electrons at the central regions are associated with those ``partially behave as mobile metallic electron".
The viewpoint of the stress tensor is closer to one of QTAIM in a sense that Li$_2$ already exhibits some aspects 
of chemical bonding which characterizes bulk Li. 

%%%%%%%%%%%%%%%%%%%%%%%%%%%%%%%
\subsection{Comparison with other molecules} \label{sec:comparison}
%%%%%%%%%%%%%%%%%%%%%%%%%%%%%%%
In this section, we analyze hydrocarbon molecules and compare with the Li clusters,
especially regarding the negativity and degeneracy of the eigenvalues of the electronic stress tensor.
Ref.~\cite{Szarek2008} has studied hydrocarbon molecules using the electronic stress tensor.
We use similar set of molecules but not exactly the same one.
The complete list of structures of our set which consists of 48 molecules is shown in the supplementary material 
(Fig.~S1 and Table S1 \cite{SI1}). 
The structures of these hydrocarbon molecules are obtained by optimization using the density functional theory (DFT) method with 
B3LYP \cite{Lee1988,Becke1993}
as a functional and the 6-311++G** basis set.
	
As for the electronic stress tensor analysis, we repeat the same analyses which has been carried out for the Li clusters
in the previous section.
The detailed data table  for the hydrocarbon molecules like Table \ref{tab:Licluster} can be found in the supplementary material (Table S2 \cite{SI1}). 
 We begin by investigating the largest eigenvalue of stress tensor.
 We find that most of the bonds have positive largest eigenvalues.
 Negative largest eigenvalues are found in some of the triple C-C bonds,
 as is mentioned in Sec.~\ref{sec:intro}.
 In our data set, the triple C-C bonds in C$_2$HX with X being H, CH$_3$, C$_2$H$_5$, NH$_2$, OH, F, Cl and Br have
 the largest eigenvalues which are negative and whose absolute values are ${\cal O}(10^{-3})$ or less
 (the three eigenvalues for these bonds against the bond lengths are plotted in Fig.~S2 in the supplementary material \cite{SI1}).
It is known that this negativity of the largest eigenvalue in these triple bonds can be attributed to the very short bond 
and not to metallicity \cite{Tachibana2005,Szarek2008,Ichikawa2011b}. 
The pattern of degeneracy of three eigenvalues of the electronic stress tensor is very different from 
that of the Li clusters. 
In the hydrocarbon molecules, the smaller two eigenvalues are degenerate, $\lambda_1 \approx \lambda_2$,
 and $\lambda_3$ has much larger values which are mostly positive. 
 Thus, if we only focus on the largest eigenvalue, these triple C-C bonds and the bonds in the Li clusters look similar.
 However, by looking at the other two eigenvalues, we can clearly distinguish them. 
 
As is done in the previous section, we compute the differential eigenvalues $\lambda_{D32}$ and $\lambda_{D21}$ at the Lagrange points
in the hydrocarbons. 
This is plotted in Fig.~\ref{fig:diff_eig} (b) and we see that $\lambda_{D32}$ is much larger than $\lambda_{D21}$.
It means that the smaller two eigenvalues are relatively degenerate and the largest one is much larger than those two.
This distribution pattern is quite different from that of the Li clusters shown in Fig.~\ref{fig:diff_eig} (a).
Let us summarize our results using the differential eigenvalues.
The Li clusters are characterized by  $\lambda_{D32} \ll 1$ and $\lambda_{D21} \ll 1$ whereas
the hydrocarbon molecules $\lambda_{D32} \gg \lambda_{D21}$ and $\lambda_{D21} \ll 1$.
In particular, $\lambda_{D32}({\rm Li}_n) \approx \lambda_{D21}({\rm Li}_n) \ll \lambda_{D32}({\rm h/c})$.
This means that the electronic stress tensor between atom pairs in the hydrocarbon molecules 
shows directionality while it does not show much directionality in the case of the Li clusters. 

In summary of this section, first of all, we find that most of the hydrocarbon molecules are
different from the Li clusters in a sense that the largest eigenvalue of the stress tensor is positive. 
Although some of the triple C-C bonds have the largest eigenvalue which is negative, they do not
look like those of the Li clusters due to the non-degeneracy of the eigenvalues.
There is no hydrocarbon molecule which has the electronic stress tensor like that of liquid,
at least in our hydrocarbon molecule set. 
Thus, this results are not inconsistent with the idea proposed in the last section that negativity and 
degeneracy of the eigenvalues of the electronic stress tensor may characterize and quantify 
some aspects of the metallicity of chemical bonding.

%%%%%%%%%%%%%%%%%%%%%%%%%%%%
\section{Conclusion} \label{sec:conclusion}
%%%%%%%%%%%%%%%%%%%%%%%%%%%%

In this paper, we have studied the electronic structure of small lithium clusters Li$_n$ ($n=2\sim 8$) using
the electronic stress tensor. 
We have found that the degeneracy pattern of the three eigenvalues $\lambda_i$ $(i=1,2,3)$ 
($\lambda_3 \ge \lambda_2 \ge \lambda_1$) of the electronic stress tensor at the Lagrange points
of the Li clusters is very much different from that of the hydrocarbon molecules,
which are covalently bonded. 
Namely, the three eigenvalues of the Li clusters have almost same values while 
the hydrocarbon molecules have the largest eigenvalue much larger than the second largest eigenvalue,
which has similar value to the smallest eigenvalue. 
The former degeneracy pattern indicates that the bonds are not directional
while the latter indicates the clear directionality of the bonds. 
The negativity of the largest eigenvalue is not unique feature of the Li clusters (some of the C-C triple bonds 
exhibit the negative largest eigenvalue) but this degeneracy is characteristic to Li.
We can consider that such difference in the degeneracy pattern reflects the metallic nature of bonding of the Li clusters 
and the covalent nature of bonding of the hydrocarbon molecules. 
Note that the negativity and degeneracy of the stress tensor, which is the property of the liquid,
is readily associated with the 
tradition that the metal can be viewed as ``electron sea".

To describe the degeneracy pattern of the eigenvalues in a compact manner, 
we have proposed to use the differential eigenvalues: 
$\lambda_{D32} = \lambda_3-\lambda_2$ and $\lambda_{D21} = \lambda_2-\lambda_1$.
By using them, the degeneracy patterns of the eigenvalues can be
summarized as 
$\lambda_{D32}({\rm Li}_n) \approx \lambda_{D21}({\rm Li}_n) \ll \lambda_{D32}({\rm h/c}) \sim {\cal O}(0.1)$.
It is of great interest whether this relation can be extended as
$\lambda_{D32}({\rm metallic}) \approx \lambda_{D21}({\rm metallic}) \ll \lambda_{D32}({\rm covalent})$
in general. 
In this paper, we have only examined the Li clusters with limited numbers of atoms.
It is important to check this relation with larger Li clusters and also with other metal clusters.
Then, the role of the electronic stress tensor in describing the metallicity would be elucidated and 
more detailed classification among the ``metallic bond" would be possible. 

\noindent 
%%%%%%%%%%%%%%%%%%%%%%%%%%%%%%%%%%%%
\section*{Acknowledgment}
%%%%%%%%%%%%%%%%%%%%%%%%%%%%%%%%%%%%
Theoretical calculations were partly performed using Research Center for
 Computational Science, Okazaki, Japan.
This work is supported partly by Grant-in-Aid for Scientific research (No.~22550011)
from the Ministry of Education, Culture, Sports, Science and Technology, Japan.

%%%%%%%%%%%%%%%%%REFERENCES%%%%%%%%%%%%%%%%%%%%%%%%%

%%%%%%%%%%%%%%%%%%%%%%%%%%%%%%%%%%%%%%%%%%%%%%%%%%%%%%%%%%%%%%%

\newpage

\begin{table}
\caption{The bond length and the eigenvalues $\lambda_i$ $(i=1,2,3)$ of the electronic stress tensor
 at the Lagrange point for H$_2$, Li$_2$ and LiH. }
\begin{center}
\begin{tabular}{|c|c|c|c|c|c|}
\hline
Bond&
length[\AA]&
$\lambda_3$&
$\lambda_2$&
$\lambda_1$
\\
\hline
\hline

${\rm H_2}$
&0.743
&0.175
&-0.219
&-0.219
\\
\hline

${\rm Li_2}$
&2.692
&-1.025$\times 10^{-3}$
&-1.025$\times 10^{-3}$
&-1.262$\times 10^{-3}$
\\
\hline

${\rm LiH}$
&1.601
&-4.954$\times 10^{-3}$
&-1.347$\times 10^{-2}$
&-1.347$\times 10^{-2}$
\\

\hline
\end{tabular}
\end{center}
\label{tab:Li2H2LiH}
\end{table}

\begin{table}
\caption{The bond length and the eigenvalues $\lambda_i$ $(i=1,2,3)$ of the electronic stress tensor
 at the Lagrange point for bonds in the Li clusters.
The numbers in the second column correspond to those labelled in Fig.~\ref{fig:structure_Licluster}. }

\begin{center}
\begin{tabular}{| l |c|c|c|c|c|c|c|}
\hline
Cluster & bond & length [{\rm \AA}] & $\lambda_3 (\times 10^3)$ & $\lambda_2 (\times 10^3)$ & $\lambda_1 (\times 10^3)$   \\
\hline
\hline
 Li$_3$
 & 1-2 & 2.782 & -0.900 & -0.993 & -1.054  \\
 & 1-3 & 3.275 & -0.271 & -0.463 & -0.781  \\
\hline

 Li$_4$
 & 1-2 & 2.667 & -0.793 & -0.919 & -1.240  \\
 & 1-3 & 3.010 & -0.498 & -0.763 & -0.981  \\
\hline

Li$_5$
 & 1-2 & 2.929 & -0.516 & -0.708 & -1.152 \\
 & 1-3 & 2.927 & -0.519 & -0.715 & -1.163  \\
 & 1-4 & 3.085 & -0.494 & -0.746 & -0.836 \\
 & 2-3 & 2.618 & -1.139 & -1.157 & -1.348 \\
 & 2-4 & 3.028 & -0.494 & -0.789 & -1.016  \\
 & 3-4 & 3.027 & -0.494 & -0.790 & -1.016  \\
\hline

Li$_6$
 & 1-2 & 2.640 & -1.004 & -1.410 & -1.410  \\
 & 1-3 & 2.849 & -0.600 & -0.924 & -1.198  \\
\hline

Li$_7$
 & 1-2 & 2.820 & -0.684 & -1.064 & -1.064  \\
 & 1-3 & 2.974 & -0.572 & -0.861 & -0.861  \\
 & 3-4 & 3.078 & -0.465 & -0.636 & -0.787  \\
\hline

 Li$_8$
 & 1-2 & 2.926 & -0.584 & -0.826 & -0.918 \\
 & 1-5 & 3.040 & -0.474 & -0.719 & -0.814 \\
\hline

\end{tabular}
\end{center}
\label{tab:Licluster}

\end{table}

\begin{table}
\caption{The bond length and the eigenvalues $\lambda_i$ $(i=1,2,3)$ of the electronic stress tensor
 at the Lagrange point for bonds in the Na clusters.
 Since Na$_3$ and Na$_4$ respectively have $C_{2v}$ and $D_{2h}$ symmetry groups, 
which are same as the Li cluster case, one may refer to Fig.~\ref{fig:structure_Licluster}
for the numbers in the second column.
Note that we do not find the Lagrange point between the atoms 1 and 3 (with the separation of 4.371\,\AA)  of Na$_3$.
}

\begin{center}
\begin{tabular}{| l |c|c|c|c|c|c|c|}
\hline
Cluster & bond & length [{\rm \AA}] & $\lambda_3 (\times 10^3)$ & $\lambda_2 (\times 10^3)$ & $\lambda_1 (\times 10^3)$   \\
\hline
\hline
 Na$_2$
 & 1-2 & 3.162 & -0.532 & -0.532 & -0.752  \\
\hline
Na$_3$
 & 1-2 & 3.316 & -0.430 & -0.444 & -0.475  \\
\hline
Na$_4$
 & 1-2 & 3.232 & -0.418 & -0.453 & -0.514  \\
 & 1-3 & 3.578 & -0.264 & -0.334 & -0.424  \\
\hline

\end{tabular}
\end{center}
\label{tab:Nacluster}

\end{table}

%%%%%%%%%%%%%%%%%%%%%%%%%%%%%%%%%%%%%%%%%%%%%%%%%%%%%%%%%%%%%%%
\ifFIG

\clearpage

%%%%%%%%%%%%%%%%%%%%%%%%%%%%%%%%%%%%%%%%%%%%%%%%%%%%%%%%%%%%%%%
%  Li2, H2, LiH varying distances
%%%%%%%%%%%%%%%%%%%%%%%%%%%%%%%%%%%%%%%%%%%%%%%%%%%%%%%%%%%%%%%
\begin{figure}
\begin{center}
\includegraphics[width=13cm]{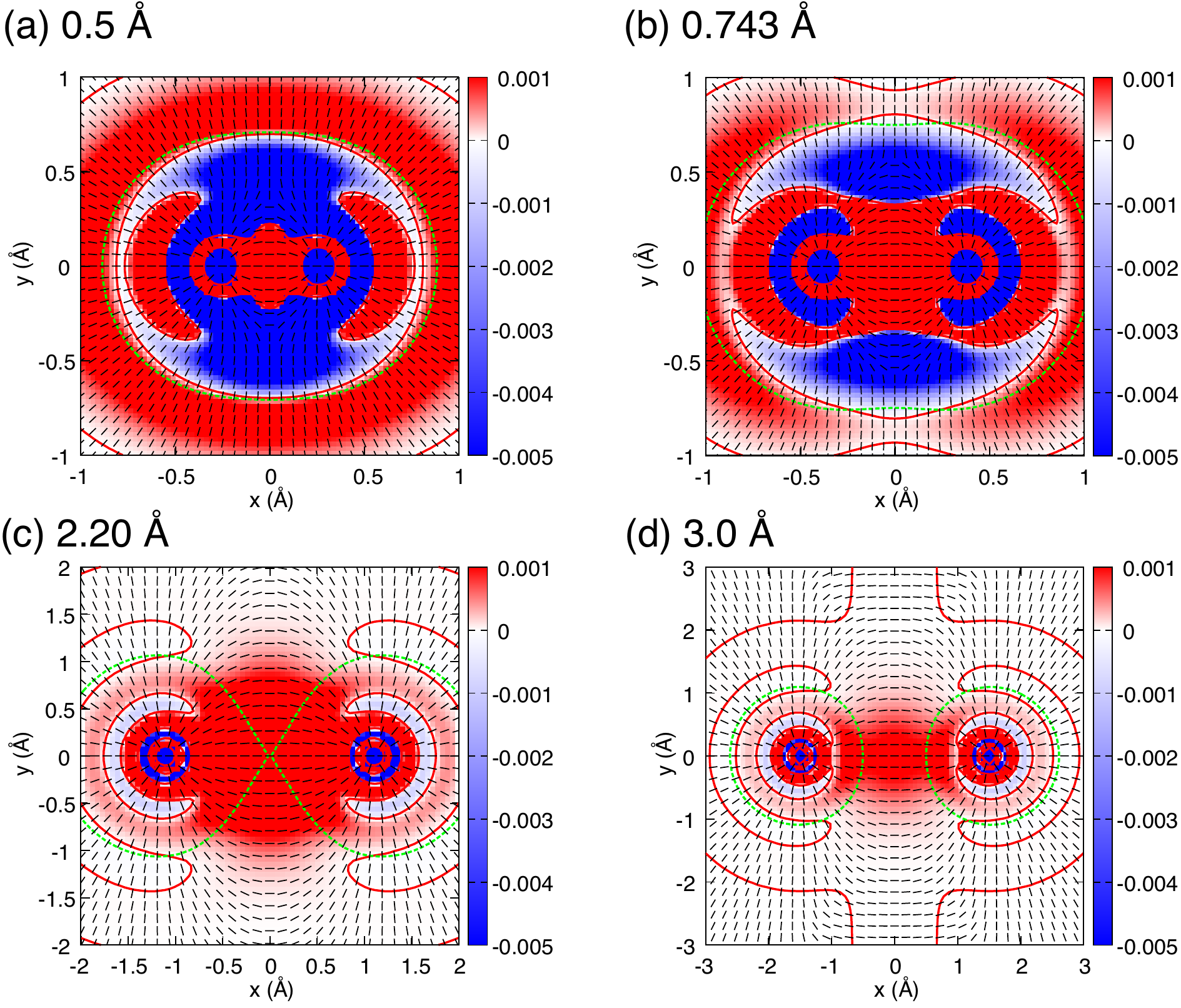}
\caption{The largest eigenvalue of the stress tensor (color map) and corresponding eigenvector (black rods)
of H$_2$ at various internuclear distances:
(a) 0.5\,${\rm \AA}$, (b) 0.743\,${\rm \AA}$ (equilibrium distance), 
(c) 2.20\,${\rm \AA}$ (intrinsic electronic transition state) and (d) 3.0\,${\rm \AA}$.
As for the eigenvectors, the projection on this plane is plotted. 
The red solid line denotes a contour where the eigenvalue is zero.
The green dashed line denotes a contour where the kinetic energy density is zero (electronic interface).
}		
\label{fig:dist_H2}
\end{center}
\end{figure}

\begin{figure}
\begin{center}
\includegraphics[width=13cm]{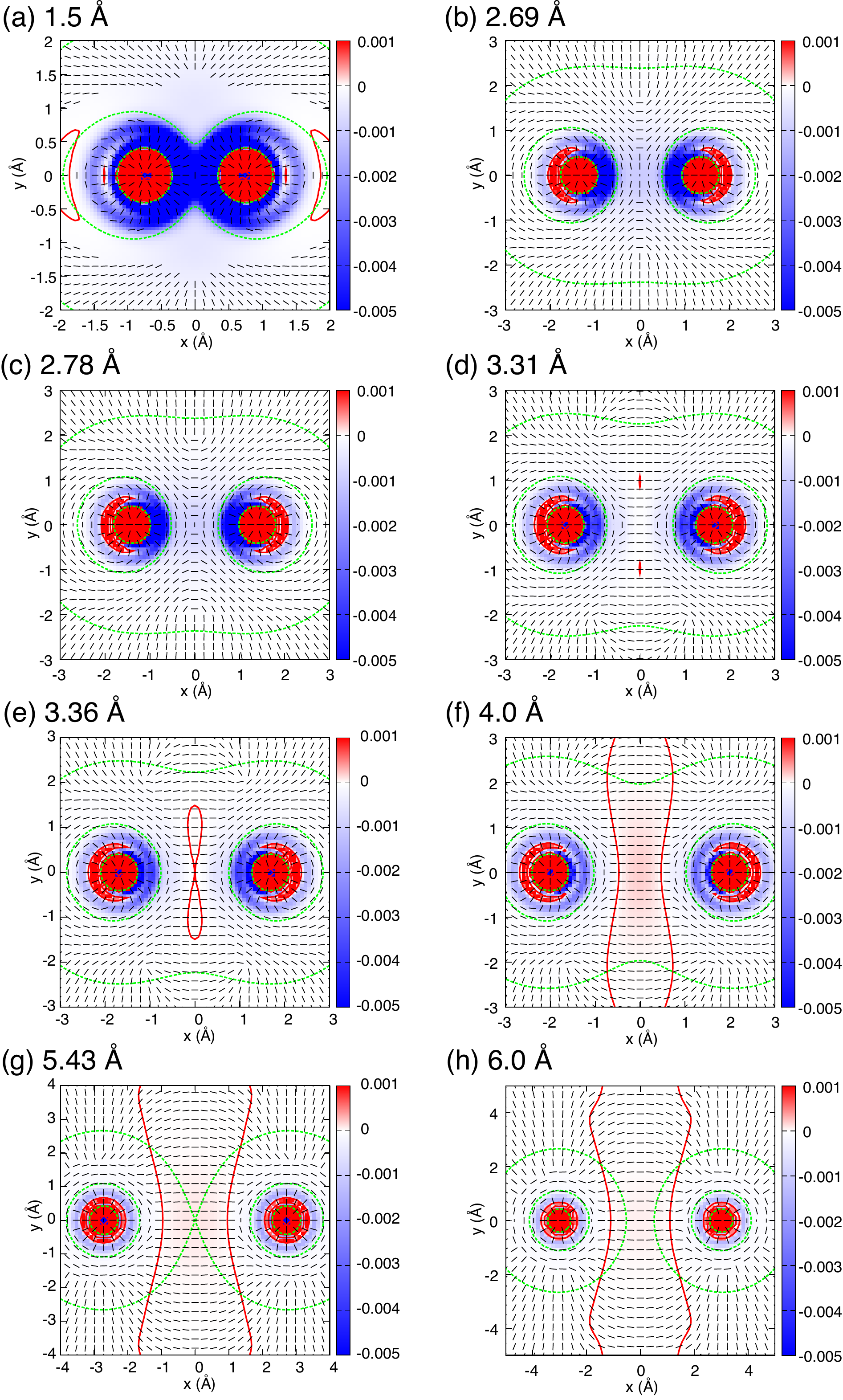}
\caption{The largest eigenvalue of the stress tensor and corresponding eigenvector
of Li$_2$ at various internuclear distances
(a) 1.5\,${\rm \AA}$, (b) 2.69\,${\rm \AA}$ (equilibrium distance), 
(c) 2.78\,${\rm \AA}$, %($\lambda_1=\lambda_2=\lambda_3$)
(d) 3.31\,${\rm \AA}$, (e) 3.36\,${\rm \AA}$, (f) 4.0\,${\rm \AA}$, 
(g) 5.43\,${\rm \AA}$ (intrinsic electronic transition state) and (h) 6.0\,${\rm \AA}$.
Plotted in the same manner as Fig.~\ref{fig:dist_H2}.
}		
\label{fig:dist_Li2}
\end{center}
\end{figure}

\begin{figure}
\begin{center}
\includegraphics[width=13cm]{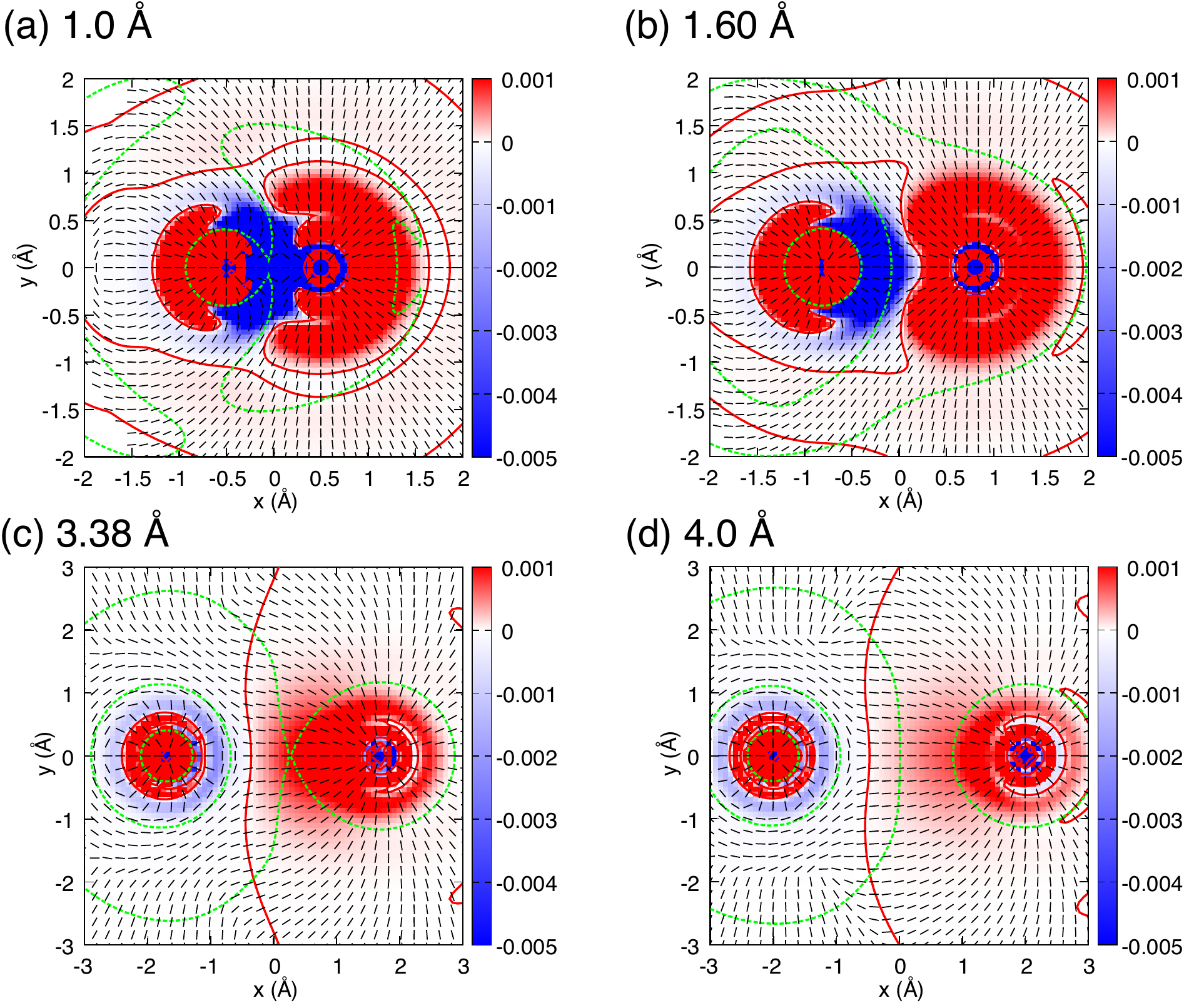}
\caption{The largest eigenvalue of the stress tensor and corresponding eigenvector of
 LiH at various internuclear distances:
(a) 1.0\,${\rm \AA}$, (b) 1.61\,${\rm \AA}$ (equilibrium distance),
 (c) 3.38\,${\rm \AA}$ (intrinsic electronic transition state) and (d) 4.0\,${\rm \AA}$.
 Plotted in the same manner as Fig.~\ref{fig:dist_H2}.
}		
\label{fig:dist_LiH}
\end{center}
\end{figure}

%%%%%%%%%%%%%%%%%%%%%%%%%%%%%%%%%%%%%%%%%%%%%%%%%%%%%%%%%%%%%%%
%  Li2, H2, LiH eig on bond axis 
%%%%%%%%%%%%%%%%%%%%%%%%%%%%%%%%%%%%%%%%%%%%%%%%%%%%%%%%%%%%%%%

\begin{figure}
\begin{center}
\includegraphics[width=8cm]{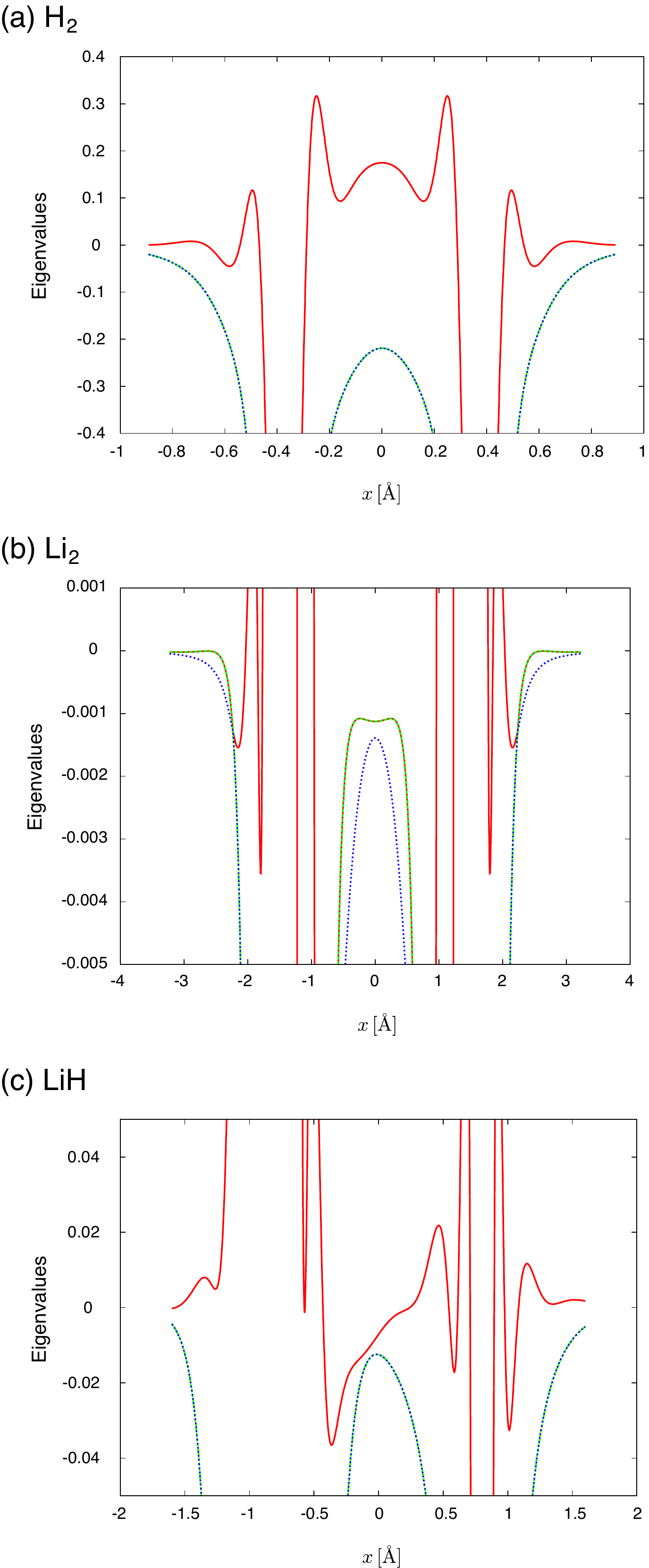}
\caption{Three eigenvalues of the stress tensor for 
H$_2$ (panel (a)), Li$_2$ (panel (b)) and LiH (panel (c)) along the internuclear axis. 
The red solid lines are for the largest eigenvalue ($\lambda_3$),
the green dashed lines are for the second largest eigenvalue ($\lambda_2$) and
the blue dotted lines are for the smallest eigenvalue ($\lambda_1$).
The internuclear distances are set to be their equilibrium distances. 
The origin of the abscissa corresponds to the midpoint of two nuclei.
}		
\label{fig:eig_Li2H2LiH}
\end{center}
\end{figure}

\begin{figure}
\begin{center}
\includegraphics[width=14cm]{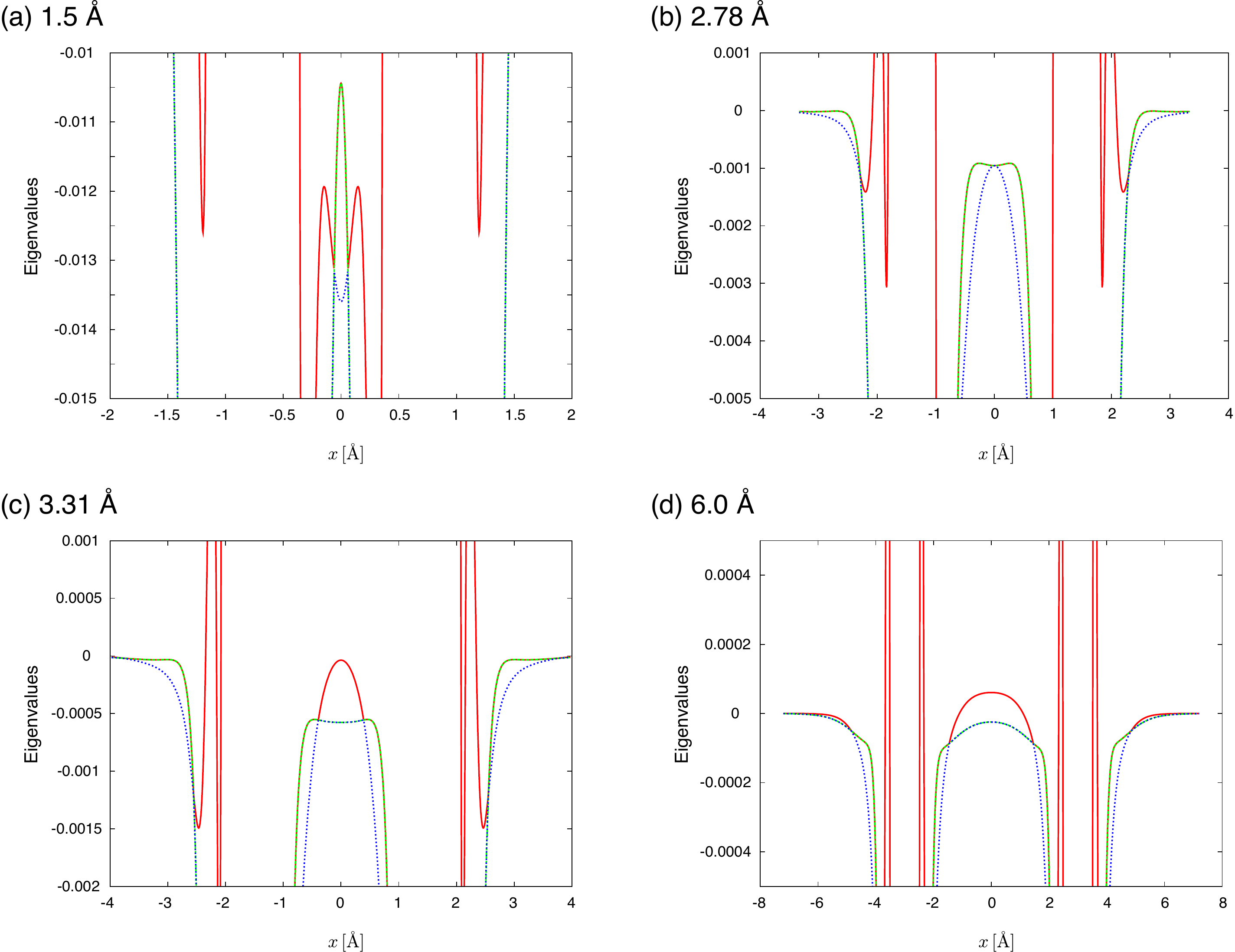}
\caption{Three eigenvalues of the stress tensor for 
Li$_2$ along the internuclear axis at various internuclear distances:
(a) 1.5\,${\rm \AA}$, 
(b) 2.78\,${\rm \AA}$, (c) 3.31\,${\rm \AA}$
 and (d) 6.0\,${\rm \AA}$.
The red solid lines are for the largest eigenvalue ($\lambda_3$),
the green dashed lines are for the second largest eigenvalue ($\lambda_2$) and
the blue dotted lines are for the smallest eigenvalue ($\lambda_1$).
}		
\label{fig:dist_Li2eig}
\end{center}
\end{figure}

%%%%%%%%%%%%%%%%%%%%%%%%%%%%%%%%%%%%%%%%%%%%%%%%%%%%%%%%%%%%%%%
%  Li clusters, bond order,  eigvec & eig on bond axis at equilibrium distance
%%%%%%%%%%%%%%%%%%%%%%%%%%%%%%%%%%%%%%%%%%%%%%%%%%%%%%%%%%%%%%%
\begin{figure}
\begin{center}
\includegraphics[width=13cm]{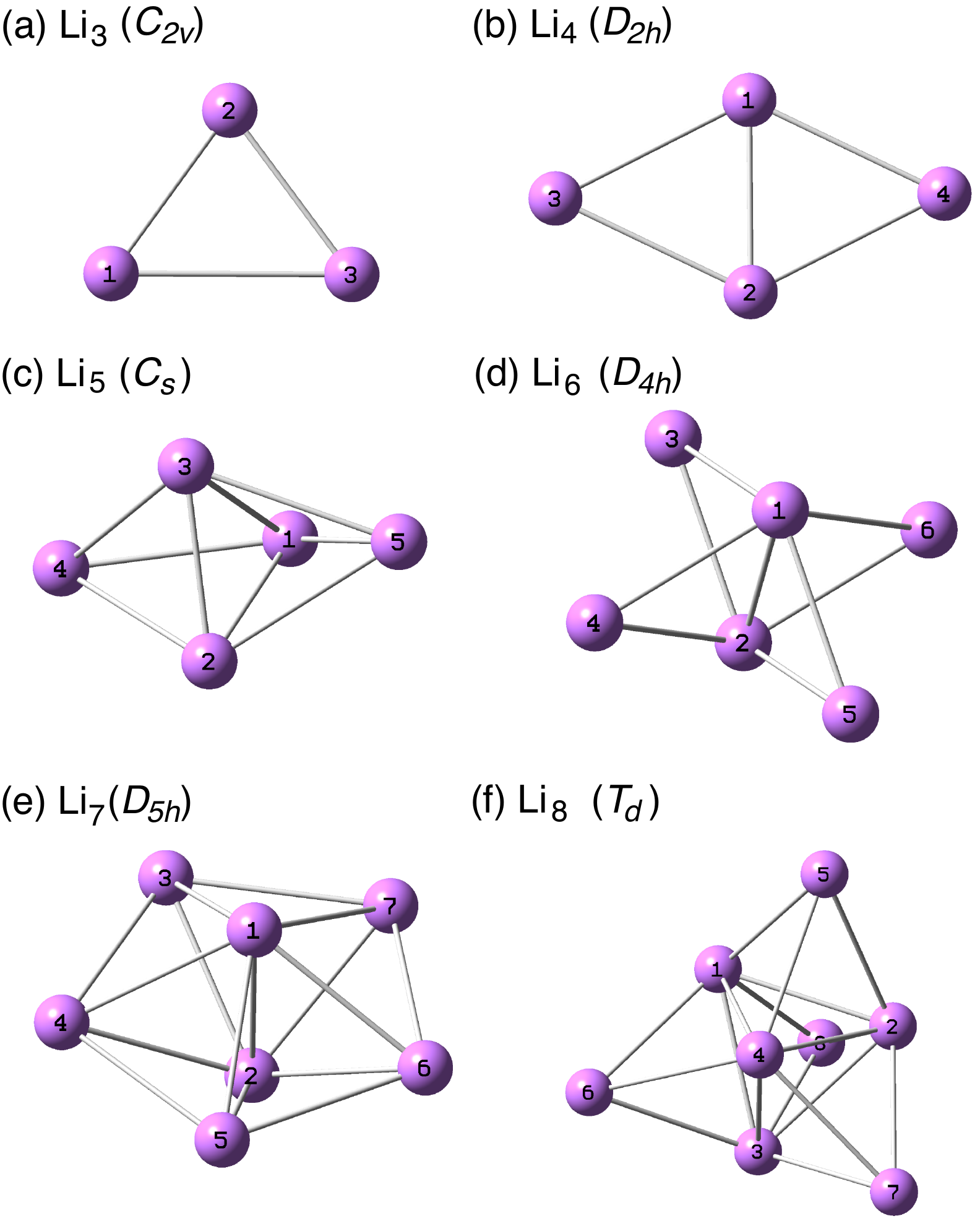}
\caption{Optimized structures of Li clusters:
(a) Li$_3$ ($C_{2v}$), (b) Li$_4$ ($D_{2h}$), (c) Li$_5$ ($C_{s}$),
 (d) Li$_6$  ($D_{4h}$), (e) Li$_7$  ($D_{5h}$) and  (f) Li$_8$  ($T_{d}$).
The bonds are drawn at which the Lagrange points are found.
}		
\label{fig:structure_Licluster}
\end{center}
\end{figure}

\begin{figure}
\begin{center}
\includegraphics[width=14.5cm]{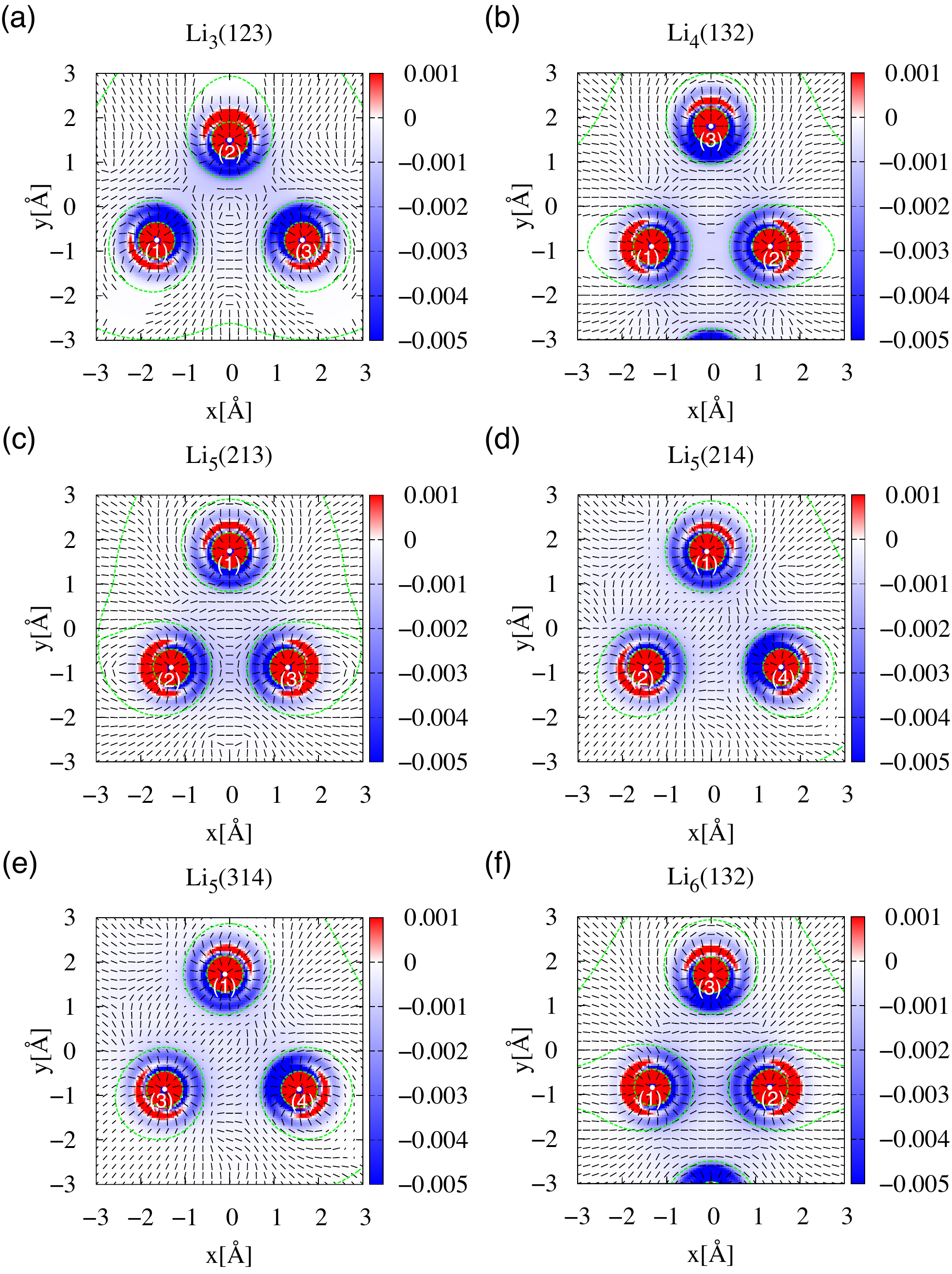}
\caption{The largest eigenvalue of the stress tensor (color map) and corresponding eigenvector for 
Li clusters Li$_n$ $(n=3\sim6)$. 
Plotted in similar manner to Fig.~\ref{fig:dist_H2}.
The filled circles denote the positions of the atoms and the numbers in parentheses correspond to
those labelled in Fig.~\ref{fig:structure_Licluster}.
}		
\label{fig:eigvec_Licluster1}
\end{center}
\end{figure}

\begin{figure}
\begin{center}
\includegraphics[width=14.5cm]{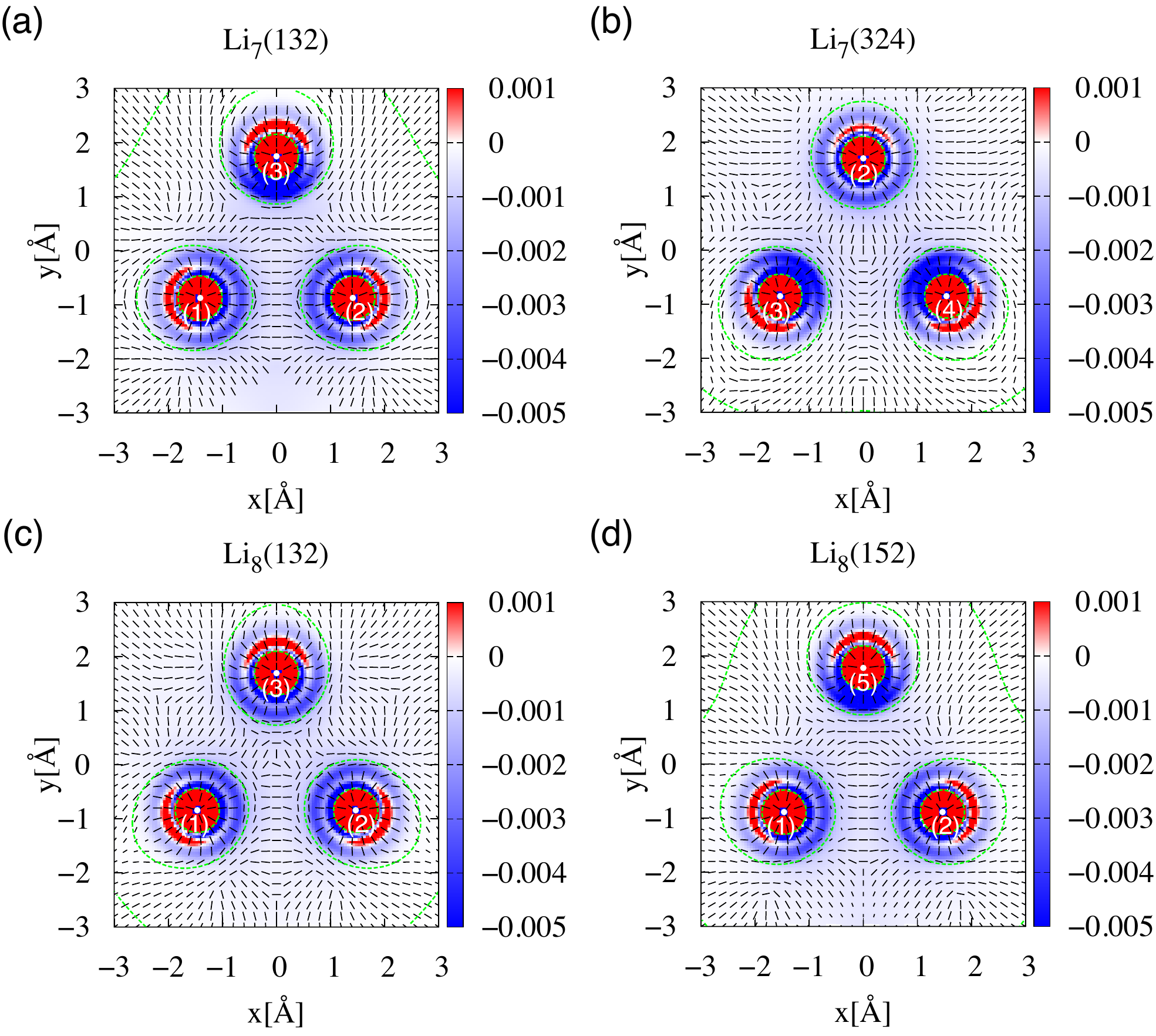}
\caption{The largest eigenvalue of the stress tensor (color map) and corresponding eigenvector for 
Li clusters Li$_n$ $(n=7\sim8)$. The optimized structures are used. 
Plotted in similar manner to Fig.~\ref{fig:eigvec_Licluster1}.
}		
\label{fig:eigvec_Licluster2}
\end{center}
\end{figure}

\begin{figure}
\begin{center}
\includegraphics[width=14cm]{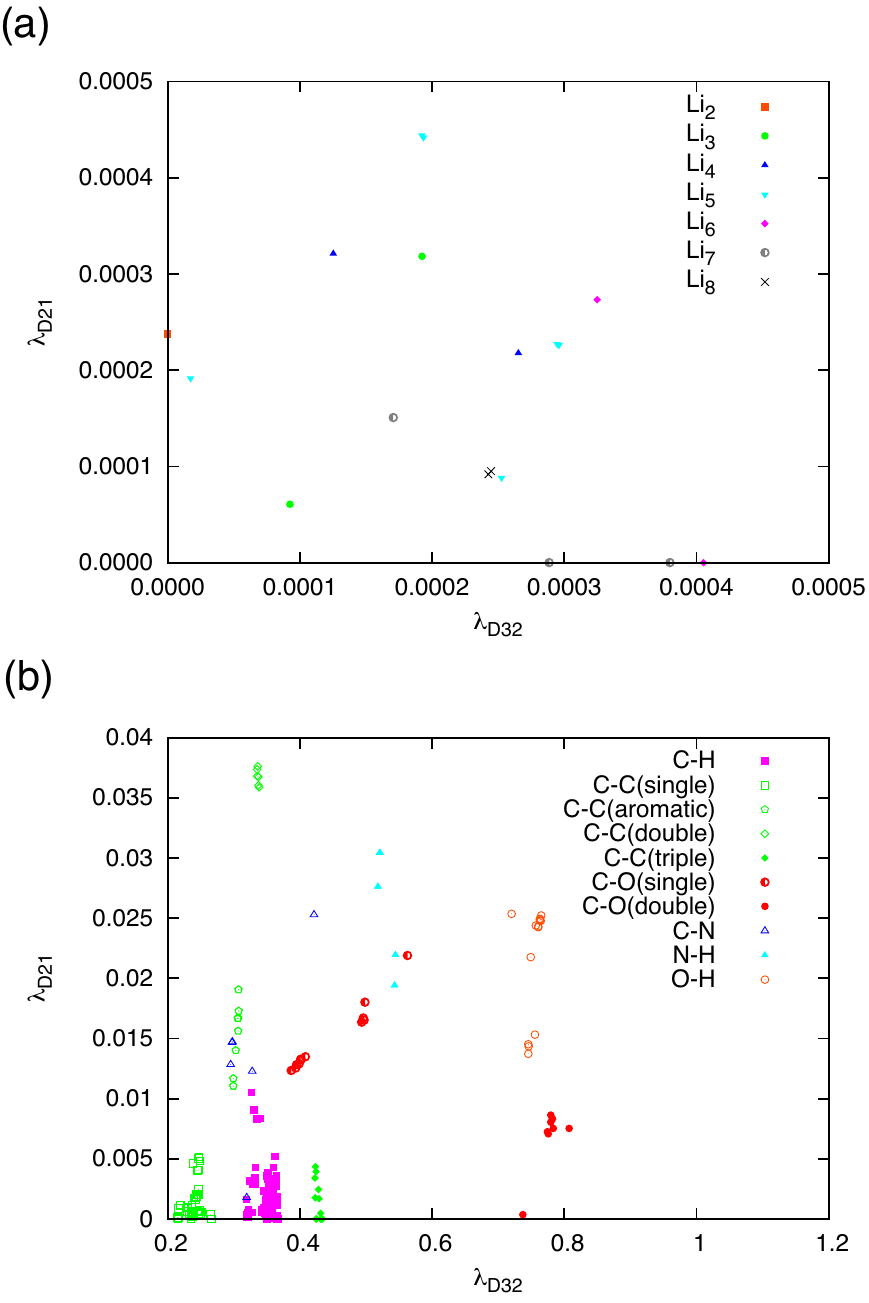}
\caption{
The relation between the differential eigenvalues, $\lambda_{D32} = \lambda_3-\lambda_2$ and $\lambda_{D21} = \lambda_2-\lambda_1$,
at the Lagrange points in the Li clusters (panel (a)) and hydrocarbon molecules (panel (b)).
 Note that the H$_2$ molecule has $\lambda_{D32} = 0.394$ and $\lambda_{D21} = 0$.
}		
\label{fig:diff_eig}
\end{center}
\end{figure}

\fi


\begin{thebibliography}{99}

\bibitem{Schrodinger1927}
	E.~Schr\"{o}dinger, Ann. Phys. (Leipzig) {\bf 82}, 265 (1927).

\bibitem{Pauli}
W. Pauli, Handbuch der Physik, Band XXIV, Teil 1, Springer, Berlin,
1933, pp.83-272; reprinted in Handbuch der Physik, vol. 5, Springer, 
Berlin,
1958. (Part 1); translated into English in General Principles of Quantum
Mechanics, Berlin, Springer, 1980.

\bibitem{Epstein1975}
	S.~T.~Epstein, J. Chem. Phys. {\bf 63}, 3573 (1975).

\bibitem{Bader1980}
	R.~F.~W.~Bader, J. Chem. Phys. {\bf 73}, 2871 (1980).

\bibitem{Bamzai1981a}
	A.~S.~Bamzai and B.~M.~Deb, Rev. Mod. Phys. {\bf 53}, 95 (1981).
	
\bibitem{Nielsen1983}
	O.~H.~Nielsen and R.~M.~Martin, Phys. Rev. Lett. {\bf 50}, 697 (1983).

\bibitem{Nielsen1985}
	O.~H.~Nielsen and R.~M.~Martin, Phys. Rev. B {\bf 32}, 3780 (1985).

\bibitem{Folland1986a}
	N.~O.~Folland, Phys. Rev. B {\bf 34}, 8296 (1986).

\bibitem{Folland1986b}
	N.~O.~Folland, Phys. Rev. B {\bf 34}, 8305 (1986).
		
\bibitem{Godfrey1988}
	M.~J.~Godfrey, Phys. Rev. B {\bf 37}, 10176 (1988).
	
\bibitem{Filippetti2000}
	A.~Filippetti and V.~Fiorentini, Phys. Rev. B {\bf 61}, 8433 (2000).

\bibitem{Tachibana2001}
	A.~Tachibana, J. Chem. Phys. {\bf 115}, 3497 (2001).

\bibitem{Pendas2002}
	A.~M.~Pend\'{a}s, J. Chem. Phys. {\bf 117}, 965 (2002).

\bibitem{Rogers2002}
	C.~L.~Rogers and A.~M.~Rappe, Phys. Rev. B {\bf 65} 224117 (2002).

\bibitem{Tachibana2004}
	A.~Tachibana, Int. J. Quantum Chem. {\bf 100}, 981 (2004).

\bibitem{Tachibana2005}
	A.~Tachibana, J. Mol. Model. {\bf 11}, 301 (2005).

\bibitem{Morante2006}
	S.~Morante, G.~C.~Rossi and M.~Testa, J. Chem. Phys. {\bf 125}, 034101 (2006).

\bibitem{Tao2008}
	J.~Tao, G.~Vignale and I.~V.~Tokatly, Phys. Rev. Lett. {\bf 100}, 206405 (2008).

\bibitem{Ayers2009}
	P.~W.~Ayers and S.~Jenkins, J. Chem. Phys. {\bf 130}, 154104 (2009).

\bibitem{Tachibana2010}
	A.~Tachibana, J. Mol. Struct. (THEOCHEM), {\bf 943}, 138 (2010).

\bibitem{Jenkins2011}
	S.~Jenkins, S.~R.~Kirk, A.~Guevara-Garc\'{i}a, P.~W.~Ayers, E.~Echegaray and A.~Toro-Labbe,
	Chem. Phys. Lett. {\bf 510}, 18 (2011).

\bibitem{Tachibana2012}
	A.~Tachibana, Frontiers in Theoretical Chemistry: Concepts and Methods: A tribute to Professor B.~M.~Deb; Eds. by Swapan K.~Ghosh and Pratim K.~Chattaraj;
Taylor \& Francis / CRC Press (2012), in press.


\bibitem{GuevaraGarcia2011}
	A.~Guevara-Garc\'{i}a, E.~Echegaray, A.~Toro-Labbe, S.~Jenkins, S.~R.~Kirk and  P.~W.~Ayers, 
	J. Chem. Phys. {\bf 134}, 234106 (2011).

\bibitem{Szarek2007}
	P.~Szarek and A.~Tachibana, J. Mol. Model. {\bf 13}, 651 (2007).

\bibitem{Szarek2008}
	P.~Szarek, Y.~Sueda and A.~Tachibana, J. Chem. Phys. {\bf 129}, 094102 (2008).

\bibitem{Szarek2009}
	P.~Szarek, K.~Urakami, C.~Zhou, H.~Cheng and A.~Tachibana, J. Chem. Phys. {\bf 130}, 084111 (2009).

\bibitem{Ichikawa2009a}
	K.~Ichikawa, T.~Myoraku, A.~Fukushima, Y.~Ishihara, R.~Isaki, T.~Takeguchi and A.~Tachibana,
	J. Mol. Struct. (THEOCHEM) {\bf 915}, 1 (2009).
	
\bibitem{Ichikawa2009b}
	K.~Ichikawa and A.~Tachibana, Phys. Rev. A {\bf 80}, 062507 (2009).

\bibitem{Ichikawa2010}
	K.~Ichikawa, A.~Wagatsuma, M.~Kusumoto and A.~Tachibana, J. Mol. Struct. (THEOCHEM), {\bf 951}, 49 (2010).

\bibitem{Ichikawa2011a}
	K.~Ichikawa, Y.~Ikeda, A.~Wagatsuma, K.~Watanabe, P.~Szarek and A.~Tachibana, 
	Int. J. Quant. Chem. {\bf 111}, 3548 (2011).

\bibitem{Ichikawa2011b}
	K.~Ichikawa, A.~Wagatsuma, Y.~I.~Kurokawa, S.~Sakaki and A.~Tachibana, 
	Theor. Chem. Acc. {\bf 130}, 237 (2011).

\bibitem{Ichikawa2011c}
	K.~Ichikawa, A.~Wagatsuma, P.~Szarek, C.~Zhou, H.~Cheng and A.~Tachibana, 
	Theor. Chem. Acc. {\bf 130}, 531 (2011).

%\bibitem{Tachibana1999}
%	A.~Tachibana, Theor. Chem. Acc. {\bf 102}, 188 (1999).

%%% Li cluster %%%%%%%%%%%%%%%%%%
\bibitem{McAdon1985}
	M.~A.~McAdon and W.~A.~Goddard III, 
	Phys. Rev. Lett. {\bf 55}, 2563 (1985).

\bibitem{Rousseau2000}
	R.~Rousseau and D.~Marx,
	Chem. Eur. J. {\bf 6}, 2982 (2000).

\bibitem{Alikhani2006}
	M.~E.~Alikhani and S.~Shaik,
	Theor. Chem. Acc. {\bf 116}, 390 (2006).
	
\bibitem{Gatti1987}
	C.~Gatti, P.~Fantucci and G.~Pacchioni,
	Theor. Chim. Acta. {\bf 72}, 433 (1987).

\bibitem{Bersuker1993}
	G.~I.~Bersuker, C.~Peng and J.~E.~Boggs,
	J. Phys. Chem. {\bf 97}, 9323 (1993).

\bibitem{Yepes2012}
	D.~Yepes, S.~R.~Kirk, S.~Jenkins and A.~Restrepo,
	J. Mol. Model {\bf 18}, 4171 (2012).
%%%%%%%%%%%%%%%%%%%%%%%%%%%%%%%


\bibitem{Gaussian09}
Gaussian 09, Revision A.1, M. J. Frisch, G. W. Trucks, H. B. Schlegel, G. E. Scuseria, M. A. Robb, J. R. Cheeseman, G. Scalmani, V. Barone, B. Mennucci, G. A. Petersson, H. Nakatsuji, M. Caricato, X. Li, H. P. Hratchian, A. F. Izmaylov, J. Bloino, G. Zheng, J. L. Sonnenberg, M. Hada, M. Ehara, K. Toyota, R. Fukuda, J. Hasegawa, M. Ishida, T. Nakajima, Y. Honda, O. Kitao, H. Nakai, T. Vreven, J. A. Montgomery, Jr., J. E. Peralta, F. Ogliaro, M. Bearpark, J. J. Heyd, E. Brothers, K. N. Kudin, V. N. Staroverov, R. Kobayashi, J. Normand, K. Raghavachari, A. Rendell, J. C. Burant, S. S. Iyengar, J. Tomasi, M. Cossi, N. Rega, J. M. Millam, M. Klene, J. E. Knox, J. B. Cross, V. Bakken, C. Adamo, J. Jaramillo, R. Gomperts, R. E. Stratmann, O. Yazyev, A. J. Austin, R. Cammi, C. Pomelli, J. W. Ochterski, R. L. Martin, K. Morokuma, V. G. Zakrzewski, G. A. Voth, P. Salvador, J. J. Dannenberg, S. Dapprich, A. D. Daniels, \"{O}. Farkas, J. B. Foresman, J. V. Ortiz, J. Cioslowski, and D. J. Fox, Gaussian, Inc., Wallingford CT, 2009.

\bibitem{ABINIT1}
X.~Gonze, B.~Amadon, P.~M.~Anglade, J.~M.~Beuken, F.~Bottin, P.~Boulanger, F.~Bruneval, D.~Caliste,
R.~Caracas, M.~Cote, T.~Deutsch, L.~Genovese, Ph.~Ghosez, M.~Giantomassi, S.~Goedecker,
D.~R.~Hamann, P.~Hermet, F.~Jollet, G.~Jomard, S.~Leroux, M.~Mancini, S.~Mazevet, M.~J.~T.~Oliveira,
G.~Onida, Y.~Pouillon, T.~Rangel, G.~M.~Rignanese, D.~Sangalli, R.~Shaltaf, M.~Torrent, M.~J.~Verstraete,
G.~Zerah, J.~W.~Zwanziger,
Computer Phys. Commun. {\bf 180}, 2582 (2009). 

\bibitem{ABINIT2}
X.~Gonze, G.~M.~Rignanese, M.~Verstraete, J.~M.~Beuken, Y.~Pouillon, R.~Caracas, F.~Jollet, M.~Torrent, G.~Zerah, M.~Mikami, Ph.~Ghosez, M.~Veithen, J.~Y.~Raty, V.~Olevano, F.~Bruneval, L.~Reining, R.~Godby, G.~Onida, D.~R.~Hamann, and D.~C.~Allan,  
Zeit. Kristallogr. {\bf 220}, 558 (2005). 

\bibitem{MRDFTv3}
	M.~Senami, K.~Ichikawa, K.~Doi, P.~Szarek, K.~Nakamura and A.~Tachibana, Molecular Regional DFT program package, ver. 3. Tachibana Lab, Kyoto University, Kyoto (2008).

\bibitem{PyMOL}
	W.~L.~DeLano, The PyMOL Molecular Graphics System. (2008) DeLano Scientific LLC, Palo Alto, CA, USA. {\tt http://www.pymol.org}

\bibitem{Gardet1996}
	G.~Gardet, F.~Rogemond and H.~Chermette, J. Chem. Phys. {\bf 105}, 9933 (1996).

% 6-311G, 6-311G**
\bibitem{Raghavachari80b}
	R.~Krishnan, J.~S.~Binkley, R.~Seeger and J.~A.~Pople, J. Chem. Phys. {\bf 72}, 650 (1980).

%6-31++G, 6-311++G
\bibitem{Frisch84}
	M.~J.~Frisch, J.~A.~Pople and J.~S.~Binkley, J. Chem. Phys. {\bf 80}, 3265 (1984).

\bibitem{Florez2008}
	E.~Florez and P.~Fuentealba, Int. J. Quant. Chem. {\bf 109}, 1080 (2009).

\bibitem{Solovyov2002}
	I.~A.~Solov'yov, A.~V.~Solov'yov and W.~Greiner, Phys. Rev. A {\bf 65} 053203 (2002).


\bibitem{Troullier1991}
	N.~Troullier and J.~L.~Martins, Phys. Rev. B {\bf 43}, 1993 (1991).


%PBE
\bibitem{Perdew1996}
J.~P.~Perdew, K.~Burke, and M.~Ernzerhof, Phys. Rev. Lett. {\bf 77}, 3865 (1996).


\bibitem{SI1}
See Supplementary Material Document No.xxxxxxxxx for the detailed data.



% LYP
\bibitem{Lee1988}
	C.~Lee, W.~Yang and R.~G.~Parr, Phys. Rev. B {\bf 37}, 785 (1988).

% B3
\bibitem{Becke1993}
	A.~D.~Becke, J. Chem. Phys, {\bf 98}, 5648 (1993).

\end{thebibliography}
\end{document}